\documentclass[%
 reprint,
nofootinbib,
 amsmath,amssymb,
 aps,
prd
]{revtex4-1}

\usepackage{lmodern}
\usepackage{graphicx}
\usepackage{dcolumn}
\usepackage{bm}
\usepackage{hyperref}
\hypersetup{linktocpage,colorlinks,citecolor={blue},pdfdisplaydoctitle=true,pdfpagemode=UseOutlines,bookmarksnumbered=true}
\usepackage{mathrsfs,dsfont}
\usepackage{color}

\makeatletter
\renewcommand{\@biblabel}[1]{[#1]\hfill}
\makeatother

\usepackage[normalem]{ulem}

\newcommand{\bra}[1]{\langle #1 |} 
\newcommand{\ket}[1]{| #1 \rangle } 
\newcommand{\upd}{\mathrm{d}}
\newcommand{\tr}{\mathrm{tr}}
\newcommand{\ie}[0]{\textit{i.e.} }
\newcommand{\eg}[0]{\textit{e.g.} }

\newcommand{\xb}[0]{\mathbf{x}}
\newcommand{\yb}[0]{\mathbf{y}}
\newcommand{\rb}[0]{\mathbf{r}}

\definecolor{cbl}{rgb}{0,0,1}
 
\definecolor{crd}{rgb}{1,0,0}

\begin{document}
\title{Ghirardi-Rimini-Weber model with massive flashes}
\author{Antoine Tilloy}
\email{antoine.tilloy@mpq.mpg.de}
\affiliation{Max-Planck-Institut f\"ur Quantenoptik, Hans-Kopfermann-Stra{\ss}e 1, 85748 Garching, Germany}
\date{\today}

\begin{abstract}
I introduce a modification of the Ghirardi-Rimini-Weber (GRW) model in which the flashes (or collapse space-time events) source a classical gravitational field. The resulting semi-classical theory of Newtonian gravity preserves the statistical interpretation of quantum states of matter in contrast with mean field approaches. It can be seen as a discrete version of recent proposals of consistent hybrid quantum classical theories. The model is in agreement with known experimental data and introduces new falsifiable predictions: (1) single particles do not self-interact, (2) the $1/r$ gravitational potential of Newtonian gravity is cut-off at short ($\lesssim 10^{-7}$m) distances, and (3) gravity makes spatial superpositions decohere at a rate inversely proportional to that coming from the vanilla GRW model. Together, the last two predictions make the model experimentally falsifiable for \emph{all} values of its parameters.
\end{abstract}

\maketitle

Most attempts to unify gravity and quantum theory rely on the quantization (fundamental or effective) of the former. Yet this route has proved so hard that, in the absence of experimental evidence, it may be time to wonder if it really is unavoidable. Could a classical curved space-time and quantum matter not coexist at the most fundamental level? This latter option has been historically dismissed on the ground that its naive instantiations yield crippling conceptual difficulties \cite{eppley1977}, already in the Newtonian limit. However, these difficulties are not shared by all semiclassical approaches \cite{albers2008}. Recently, new proposals have been put forward to construct consistent (non-relativistic) hybrid theories of gravity \cite{kafri2014,kafri2015,tilloy2016,tilloy2017}. 
At a formal level, these proposals implement gravity as a continuous measurement and feedback scheme on quantum matter. \emph{Equivalently}, the proposals take the form of continuous collapse models in which a noisy modification of the mass density sources the gravitational field. The first point of view ensures that the models have a transparent empirical content and the second guarantees that they have a clear ontological interpretation and solve the measurement problem as a byproduct. 

The objective of this short article is to introduce similar but conceptually more transparent and technically easier model. Starting from the simplest discrete collapse model, the Ghirardi-Rimini-Weber (GRW) model \cite{ghirardi1986}, instead of its continuous refinements \cite{diosi1989,ghirardi1990csl,bassi2013}, we will implement a gravitational interaction by making the flashes (collapse space-time events) ``massive''. The resulting model will yield a similar phenomenology while keeping technicalities to a minimum. The aim is not to propose a ``final'' model of semi-classical gravity but to demonstrate that a consistent approach, with clear physical and empirical contents, is possible. More modestly, even if fundamental semiclassical gravity is ultimately discarded, the present model may provide an effective description for a deeper quantum theory.

\section{The GRW model and its interpretation}

We consider for simplicity an isolated system of $N$ distinguishable particles of coordinates $\{\xb_1,\cdots,\xb_N\}$, masses $\{m_1,\cdots,m_N\}$, and without spin. The $N$-particle system is described by the wave-function $\psi_t(\xb_1,\cdots,\xb_n)$, with $\psi_t \in \mathrm{L}^2(\mathds{R}^{3N},\mathds{C})$ which obeys a standard Schr\"odinger equation $ \partial_t \psi_t = -\frac{i}{\hbar} \hat{H}_{0}\, \psi_t$, where $\hat{H}_{0}$ can be any many-body Hamiltonian. In the GRW model, this unitary evolution is interrupted by spontaneous jumps of the wave-function that can hit each particle independently:
\begin{equation}\label{eq:jump}
    \psi_t \longrightarrow \frac{\hat{L}_k(\xb_f) \psi_t}{\|\hat{L}_k(\xb_f) \psi_t\|},
\end{equation}
with:
\begin{equation}
\hat{L}_k(\xb_f)=\frac{1}{(\pi r^2_C)^{3/4}} \mathrm{e}^{-(\hat{\xb}_k-\xb_f)^2/(2 r^2_C)},
\end{equation}
where $\hat{\xb}_k$ is the position operator associated to particle $k$.
These $N$ types of jumps occur independently of each other and are uniformly distributed in time with intensity $\lambda$.  Finally, if a jump hits particle $k$ at time $t$, the probability density for the center of the collapse $\xb_f$ is given by $P(\xb_f)=\|\hat{L}_k(\xb_f) \psi_t\|^2$. 

This model introduces two new parameters: the width of the collapsed wave-function $r_C$ and the jump rate $\lambda$. They need to have small enough values that collapse events go unnoticed for small $N$, making the standard unitary evolution approximately valid, but large enough that macroscopic superpositions (with $N\gtrsim10^{20}$) are almost immediately collapsed in definite locations. The typical values proposed in the literature, $\lambda=10^{-16}\mathrm{s}^{-1}$ and $r_C=10^{-7}\mathrm{m}$, can be shown to achieve this goal \cite{ghirardi1986,feldmann2012} (although they might soon be experimentally falsified).

The GRW model is a quantum theory without observer in the sense that what happens in measurement situations is derived rather than postulated. 
As measurement results are no longer primitive in this approach, it raises the question: what is the theory fundamentally about? One proposal, going back to John Bell \cite{bell1987}, is to regard the localized collapse events or flashes  $(\xb_f,t_f)$ as the elementary building blocks of matter (``local beables" \cite{bell1976} or ``primitive ontology" \cite{allori2014,allori2015,tumulka2011}), of which, in particular, all macroscopic objects are composed. 
Our proposed model takes this interpretation seriously: if the flashes are matter, they should gravitate. The role of the wave-function, on the other hand, is only to provide the probabilistic law for their distribution. It does not represent matter and hence does not source a gravitational field.

An important feature of the GRW model, and of all collapse models for that matter, is that it is consistent with the statistical interpretation of the wave-function. Once one understands how the measurement postulate is \emph{derived} from the model, one can use the wave-function exactly as in orthodox quantum theory (with the Born rule and collapse in measurement situations) but with a small modification of the dynamics. This is crucial to make the theory usable in practice.

The probabilities one can compute in orthodox quantum theory are obtained as sesquilinear forms on the state $\ket{\psi}$, \ie they are quantities of the form $a = \bra{\psi} \hat{A} \ket{\psi}$, where $\hat{A}$ is a Hermitian operator. 
In collapse models, one does not have a direct empirical access to the random jumps. One can only know their macroscopic configuration a posteriori as they determine measurement outcomes. As a result, the empirical content of the model --what one can predict-- is contained in quantities of the form
$\bar{a}=\mathds{E}\left[\bra{\psi} \hat{A} \ket{\psi}\right]=\tr \left(\hat{A}\,\mathds{E}\big[ \ket{\psi}\bra{\psi}\big]\right)$,
where $\mathds{E}[\, \cdot \,]$ denotes the statistical average over the random jumps. Thus, although the non-linear random state evolution is necessary to understand why the theory makes sense, all the empirical content lies ultimately in the density matrix $\rho=\mathds{E}\big[ \ket{\psi}\bra{\psi}\big]$. A straightforward computation shows that it obeys the equation:
\begin{equation}
\partial_t \rho_t = -\frac{i}{\hbar}[\hat{H}_0,\rho_t] + \lambda\! \sum_{k=1}^N\!\int \! \upd \xb_f  \hat{L}_k(\xb_f) \rho_t \hat{L}_k(\xb_f) \! - \!\rho_t. \label{eq:master} 
\end{equation}
Importantly, equation \eqref{eq:master} is \emph{linear}. If this were not the case, the standard statistical interpretation would necessarily break down \cite{gisin1990,polchinski1991,bassi2015}. Indeed, if \eqref{eq:master} were non-linear, it would mean that proper mixtures (statistical ensembles of pure states) would no longer be empirically equivalent to improper mixtures (decohered quantum superpositions of pure states obtained by tracing out unknown degrees of freedom). This equivalence is what usually guarantees that Alice and Bob cannot communicate using an EPR pair: measuring her state, Alice cannot know what Bob did with his. With non-linearities, the statistics on Alice's side would depend on Bob's measurements. The standard objection is that it would allow faster than light signalling. The difficulty is actually even more crippling: with non-linearities, reduced states can simply no longer be trusted to make predictions. The only states one ever has access to in practice being reduced states, there can be no consistent orthodox statistical interpretation. This does not mean that the model as such would be inconsistent, but since reduced states are the only ones to which we have empirical access in practice, the usual statistical formalism of quantum mechanics could no longer be applied\footnote{The linearity of the equation for the density matrix in collapse models plays the same role as ``equivariance'' in Bohmian mechanics \cite{durr2009}. Without it, the theory might still make sense but its empirical content becomes almost impossible to extract.}.

I insist on this linearity at the master equation level because it is at the same time necessary and unexpected. After all, we started with a non-linear collapse evolution! Yet, the jump probabilities exactly simplify with the normalization factors of \eqref{eq:jump} when averaging over the noise, thus magically removing the non-linearity for the density matrix. 
Of course, there is no magic here: the GRW model has been constructed in such a way as to reproduce quantum statistics. However, an arbitrary modification of the stochastic evolution will most likely not preserve this feature. Unless there is a particular reason why linearity (of the master equation) should hold, it typically will not. Conversely, this means that the prerequisite of linearity puts tight constraints on possible extensions or modifications of the GRW model (e.g. when adding gravity).

A possible understanding of why collapse models are consistent with the statistical interpretation comes with noticing that they can be formally rewritten as purely orthodox models. The ``jump'' part of the GRW evolution \eqref{eq:jump} takes the form of a positive-operator valued measure (POVM), \ie a generalized measurement perfectly authorized in ``operational'' quantum mechanics. Averaging over measurement results is operationally equivalent to tracing out the degrees of freedom of the measurement apparatus and naturally gives a perfectly legitimate linear open system evolution \eqref{eq:master} of the Lindblad form. The interpretation of the GRW model is naturally different: no observer is fuzzily measuring the particle positions and the jumps are taken as fundamental. Yet, the fact that the formalism can be recast in orthodox terms can be seen as a guarantee that the operational toolbox of quantum theory will survive. 

In the orthodox picture, the flashes correspond to measurement outcomes. Measurement outcomes have a nice status in quantum theory: they are the only ``classical'' variables one has access to. In particular, measurement outcomes can be fed back into the evolution of the quantum state without causing trouble. Some controller can apply an external field to a quantum system contingent on a measurement outcome without wreaking havoc on the statistical interpretation of states. If it were not so, the operational toolbox of quantum theory would be unusable. Obviously in the GRW model, no observer has access to individual flashes. However, because they are equivalent to measurement outcomes at a formal level, one can make the wave-function dynamics depend on them. If we are in the business of constructing new theories, it is legitimate to use the flashes as classical sources. This makes the flashes very different from other primitive ontologies proposed in foundations. Neither the quantum expectation value of the mass density operator nor the Bohmian particle trajectories can backreact on the wave-function without ruining the statistical interpretation. To my knowledge, the flashes (or their field equivalent for continuous collapse models) are the only candidates for a primitive ontology so far proposed that need not be purely \emph{passive}\footnote{I should insist that being passive is in no way a fundamental problem for a primitive ontology. Being able to consistently back react on the wave-function simply offers more flexibility for theory building.} (\ie determined by the wave-function without influencing it in return).

\section{Sourcing gravity from the flashes}

Guided by the previous intuition, we can use the flashes to source a classical gravitational field $\Phi$ (instead of the expectation value of the mass density discussed in \cite{derakhshani2014}). We start by considering that a flash just happened and forget the rest of the dynamics. For a flash $(\xb_f,t_f)$ associated to the collapse operator $\hat{L}_k$ and thus to particle $k$, we source $\Phi$ using the Poisson equation:
\begin{equation}\label{eq:poisson}
\Delta \Phi(\xb,t) = 4\pi G m_{k} \lambda^{-1} f(t-t_f, \xb-\xb_f),
\end{equation}
where $f$ is a space-time form factor that smears the flash in space-time. It should satisfy $\int\upd^3\xb\, \upd t\, f(t,\xb)=1$, $f\geq 0$ and $\forall t<0, f(t,\xb)=0$. As the particles flash at a rate $\lambda$, the factor $\lambda^{-1}$ insures that the field generated by the flashes corresponds on average to the field created by a classical particle in the same position. We will mainly be interested in the singular limit where the gravitational potential is sourced by the sharp flash only and thus take $f(t-t_f, \xb-\xb_f) \rightarrow \delta(t-t_f) \delta^{3}(\xb-\xb_f)$. However, a general $f$ may be useful\footnote{An example of a more regular sourcing would be to take the same Gaussian space smearing as the one used in the collapse operators $\hat{L}_k(\hat{\xb})$, while keeping a Dirac in time to preserve Markovianity. This minor extension can be treated without difficulty.} for extensions out of the Newtonian limit where point sources would yield divergent metrics. 

We now add the corresponding gravitational potential in the evolution for the wave-function as a totally benign external potential:
\begin{equation}
\hat{V}_{\rm G}= \int \upd \xb \, \Phi(\xb)\, \hat{M}(\xb)
\end{equation}
where $\hat{M}(\xb) = \sum_{\ell=1}^N m_\ell \, \delta(\xb-\hat{\xb}_\ell)$ is the mass density operator. Inverting the Poisson equation \eqref{eq:poisson} gives:
\begin{equation}\label{eq:potential}
\hat{V}_G(t)= -G \lambda^{-1} \sum_{\ell=1}^N m_k m_\ell \int \upd \xb  \frac{f(t-t_f,\xb-\xb_f)}{|\xb-\hat{\xb}_\ell|} .
\end{equation}
This is the potential created by a single flash. By linearity, the total potential is simply the sum of the potentials associated with all the past flashes.
Taking the limit $f(t-t_f, \xb-\xb_f) \rightarrow \delta(t-t_f) \delta^{3}(\xb-\xb_f)$ will make the potential \eqref{eq:potential} created by each flash divergent but the corresponding unitary $\hat{U}_k(\xb_f,k) = \exp\left(-i\int_{t_f}^{+\infty} \upd t \, \hat{V}_G(t)/\hbar\right)$ applied on the wave-function --instantaneous in the limit-- will remain finite. Indeed, in that limit we get:
\begin{equation}
\hat{U}_k(\xb_f)=\exp\left(i \,\frac{G}{\lambda \hbar}\sum_{\ell=1}^N \frac{m_k m_\ell}{|\xb_f-\hat{\xb}_\ell|}\right).
\end{equation}

Hence in the absence of smearing, the physical picture becomes particularly simple. Immediately after a flash, an instantaneous unitary transformation is applied to the wave-function corresponding to the singular gravitational pull of the flash. The evolution is thus very close to that of the original GRW model but with a modification of the jump operators. Indeed, after particle $k$ flashes, the wave-function experiences the transformation:
\begin{equation}
\psi_t \longrightarrow \hat{U}_k(\xb_f)\frac{ \hat{L}_k(\xb_f)\psi_t }{\| \hat{L}_k(\xb_f)\psi_t\|} \equiv \frac{\hat{B}_k(\xb_f) \psi_t }{\| \hat{B}_k(\xb_f)\psi_t\|}.
\end{equation}
As before we obtain the master equation for the density matrix:
\begin{equation}\label{eq:master2}
\partial_t \rho_t = -\frac{i}{\hbar}[\hat{H}_0,\rho_t] + \lambda\! \sum_{k=1}^N\!\int \upd \xb_f \hat{B}_k(\xb_f)\rho_t \hat{B}^\dagger_k(\xb_f) \!- \!\rho_t.
\end{equation}
This is the central result of this short article. 

\section{Phenomenology}

\label{sec:singleparticle}
We now study the evolution of a single isolated particle. Ignoring the free Hamiltonian $\hat{H}_0$ to focus on the gravitational and collapse effects, we see that the master equation \eqref{eq:master} is diagonal in position:
\begin{equation}
\partial_t \rho_t(\xb,\yb) = \lambda(\Gamma(\xb,\yb)-1)\rho_t(\xb,\yb),
\end{equation}
where $\Gamma(\xb,\yb)$ is an \emph{a priori} complex kernel:
\begin{equation}\label{eq:kernel}
\begin{split}
\Gamma(\xb,\yb)\!=\!\!\int \!\! \frac{\upd\xb_f}{(\pi r_C^2)^{3/2}} &\exp \left(i\frac{G m^2}{\lambda \hbar}\left[\!\frac{1}{|\xb-\xb_f|}-\frac{1}{|\yb-\xb_f|}\!\right]\right)\\
\times &\exp\left(-\frac{(\xb-\xb_f)^2+(\yb-\xb_f)^2}{2 r_C^2}\right).
\end{split}
\end{equation}
Changing of variable with $\tilde{\xb}_f = \xb_f + (\xb+\yb)/2$, one sees that the integral on the upper half space is equal to the conjugate of the integral on the lower half space, hence $\Gamma$ is \emph{real}. It means that the effect of collapse and gravity on a single particle is simply a decay of the phases in the position basis. A single particle does not attract itself in this model. This absence of self-attraction is a counterintuitive prediction markedly distinguishing this approach from other semi-classical theories relying on mean field.

We further notice that the master equation \eqref{eq:master2} involves a new length scale $r_G = G m^2/(\hbar \lambda)$. For $\lambda=10^{-16}\mathrm{s}^{-1}$, we get $r_G\simeq1.8\times 10^{-14}\mathrm{m}$ for protons and $r_G \simeq 5.3\times 10^{-21} \mathrm{m}$ for electrons. This length scale appears notably in the short distance behavior of decoherence. Indeed, from \eqref{eq:kernel} we get:
\begin{equation}
    \lambda (1-\Gamma(\xb,\yb) ) \underset{|\xb-\yb|\rightarrow 0}{\sim}  \frac{\lambda}{\sqrt{\pi}} \frac{r_G^2}{r_C^{3}} \, |\xb-\yb| \propto \lambda^{-1}|\xb-\yb|.
\end{equation}
Importantly, this means that the coupling with gravity creates decoherence effects of strength inversely proportional to the collapse rate $\lambda$: the range of values allowed for the latter is experimentally falsifiable from below. 

Finally, we need to make sure that the model is at least empirically adequate in the Newtonian limit. For that matter, we consider a test particle of coordinate $\xb_0$ interacting with a lump of $N$ particles of coordinates $\xb_1,\cdots,\xb_N$. The only thing we assume is that the test particle is separated from the lump by a distance much larger than $r_C$.
Ignoring again $\hat{H}_0$, we can expand \eqref{eq:master2} to first order in $\varepsilon=r_G/r_C<10^{-7}m$ to write the approximate master equation for this set of $N+1$ particles:
\begin{equation}\label{eq:master3}
\begin{split}
&\partial_t \rho_t\simeq  \lambda\sum_{k=0}^N \left(\int \upd \xb_f \hat{L}_k(\xb_f) \rho_t \hat{L}_k(\xb_f) - \rho_t\right) \\
&+ i\!\sum_{k,\ell=0}^N\!\frac{G m_k m_\ell}{\hbar}\!\int\!\upd \xb_f\!\left[\frac{1}{|\hat{\xb}_\ell-\xb_f|}-\frac{1}{|\hat{\yb}_\ell-\xb_f|}\right]\\
&\hskip4.5cm \times\hat{L}_k(\xb_f) \rho_t  \hat{L}_k(\xb_f).
\end{split}
\end{equation}
We shall be interested in the reduced density matrix of the test particle $\rho_t^\circ(\xb_0,\yb_0) = \tr_N(\rho_t)(\xb_0,\yb_0)$.
Taking the trace in \eqref{eq:master3} cancels most of the terms and one gets:
\begin{equation}\label{eq:masterreduced}
\begin{split}
&\partial_t \rho_t^\circ(\xb_0,\yb_0) = \int \upd \xb_f \hat{L}_0(\xb_f) \rho^\circ_t \hat{L}_0(\xb_f) - \rho^\circ_t\\
&+ \!i\! \sum_{k=1}^{N} \frac{Gm_0 m_k}{\hbar}\!\! \! \int \!\!\upd \xb_1\cdots \upd \xb_N\, \upd\xb_f \!\! \left[\!\frac{1}{|\xb_0-\xb_f|}-\frac{1}{|\yb_0-\xb_f|} \!\right]\\
&\times \frac{1}{(\pi r_C^2)^{3/2}}\mathrm{e}^{-\frac{|\xb_k-\xb_f|^2}{r_C^2}} \!\!\!\rho_t(\xb_0,\xb_1,\cdots,\xb_N,\yb_0,\xb_1,\cdots,\xb_N).
\end{split}
\end{equation}
We now use the fact that the test particle is far from the lump to neglect the Gaussian convolution in \eqref{eq:masterreduced}:
\begin{equation}
\partial_t \rho_t^\circ \simeq \lambda\left(\Gamma_0 (\rho_t^\circ)-\rho_t\right) - i \,  \tr_N \left\{ \left[-\sum_{k=1}^N \frac{G m_0 m_k}{\hbar |\hat{\xb}_0-\hat{\xb}_k|},\rho_t\right] \right\},
\end{equation}
where $\Gamma_0(\rho_t^\circ)(\xb_0,\yb_0)=\Gamma_0(\xb_0,\yb_0) \rho_t^\circ(\xb_0,\yb_0)$ contains the single particle decoherence coming from the original GRW model. Equation \eqref{eq:masterreduced} shows that the statistics on the test particle are approximately \emph{as if} it were interacting with the $N$ particles of the lump through a (quantum) Newtonian pair potential. The crucial (and only) hypothesis to obtain this result is that the distance between the test particle and the lump is much larger than $r_C$.

If we further assume that the particles in the lump are well localized (typically because of the collapse mechanism itself) around positions $\rb_1,\cdots,\rb_N$ on a length scale far smaller than the lump-test particle distance, we can remove the trace and obtain the classical limit:
\begin{equation}
\partial_t \rho_t^\circ \simeq \lambda\left(\Gamma_0 (\rho_t^\circ)-\rho_t\right) - i  \left[-\sum_{k=1}^N \frac{G m_0 m_k}{\hbar \, |\hat{\xb}_0-\rb_k|},\rho^\circ_t\right].
\end{equation}
Apart again from the decoherence coming from the collapse, this is exactly what one would have wished for. A classical piece of matter approximately creates the classical gravitational field one would naively expect, at least at a distance $d \gg r_C\sim 10^{-7}$m away from it. At shorter distances, the effective Newtonian pair potential is smoothed by the collapse operators and the $1/r$ law of Newtonian gravity breaks down. However, this behavior is still compatible with all the experiments carried out so far where the gravitational field probed is typically sourced by large objects at least $10^{-4} \mathrm{m}$ away from the test mass \cite{hagedorn2015}.


Including gravity in the GRW model in the way we have suggested, \ie by making the flashes gravitate, has an interesting consequence on the parameter diagram of the model. The problem of vanilla collapse models is that parameters leading to very unsharp or very infrequent collapse can only ever be philosophically discarded. With a $\lambda$ too small or $r_C$ too large, the GRW model does not reduce macroscopic superpositions before they include different observer states. This leads, in effect, to a Many-Worlds theory that cannot be empirically discarded but undermines the main motivation for considering the collapse mechanism in the first place. However, there is no way to experimentally falsify this fuzzily defined zone of metaphysical discomfort. 

The present model is free of this issue. The leading decoherence term is $\propto \lambda$ (the ``instrinsic'' GRW decoherence) but the coupling with gravity adds decoherence $\propto \lambda^{-1}$. This means that for a fixed value of $r_C$, all the values of $\lambda$ can \emph{in principle} be experimentally falsified. Intuitively, an infrequent collapse yields infrequent but supermassive flashes and thus a noisy gravitational interaction yielding strong gravitational decoherence. Reciprocally, reducing gravitational decoherence requires frequent flashes and thus a strong intrinsic decoherence.

Further, the effective potential one obtains is smoothed by the collapse operators and the $1/r$ law of gravity breaks down for distances $d\lesssim r_C$. Hence, small values of $r_C$ can be falsified by decoherence and large values of $r_C$ can be falsified by probing the gravitational force at short distances. The model is --\emph{in principle}-- experimentally falsifiable for all values of $\lambda$ and $r_C$.

\section{Discussion}

The dynamical equations of collapse models like the GRW model are the same as the one that describe repeated unsharp measurements of particle positions (or mass density for continuous collapse models). This formal equivalence is ultimately what guarantees that the statistical interpretation of quantum states survives in such models. Further, it shows that the space-time events (or flashes) labeling collapse events are \emph{formally} measurement outcomes. This means that, even though they are not known to observers, one can let them backreact explicitly on the quantum state as one would in a feedback scheme. Such a coupling does not create the usual difficulties of mean field couplings which yield non-linearities at the master equation level \cite{bahrami2014}. This is a crucial insight for theory building allowing to consistently couple matter with a classical gravitational field. Following this idea, we made the flashes themselves source the gravitational field, as infinitely ``massive'' space-time events. The model we obtained gives the Newtonian gravitational force one would expect between particles for distances larger than the collapse length scale $r_C$ and thus is in agreement with known experiments. However, particles do not attract themselves, contrary to what one would have guessed from naive semi-classical theories. The model is falsifiable for all possible values of its two parameters $\lambda$ and $r_C$ in contrast with standard collapse models where only large values of $\lambda$ and small values of $r_C$ can be experimentally eliminated.

In the end, if one pays the price of collapse models to solve the measurement problem, the possibility to construct consistent hybrid dynamics comes for free. The implementation of such dynamics then appreciably constrains the parameter diagram of collapse models.
The present approach thus has appealing features.
That said, how much should we \emph{believe} in the model introduced here? Theoretical physics is already drowning in wild speculations, sometimes prematurely elevated to the status of truth, hence a bit of soberness is required. The present model is a proof of principle rather than a proposal for a realistic fundamental theory. Nonetheless, important lessons survive its limited scope:
\begin{enumerate}
\item There is no obstacle in principle to construct consistent semi-classical theories of gravity.
\item Collapse models can be empirically constrained by a natural coupling with gravity.
\item A primitive ontology can have a central dynamical role and need not be merely passive. 
\end{enumerate}
If semi-classical theories of the type presented here can be extended to general relativity\footnote{On top of the difficulty to construct relativistic collapse models, one may worry that no divergence-free quantity $T_{\mu\nu}$ (needed to source gravity in general) can be constructed from a flash-like object. This latter problem might be bypassed by going to unimodular gravity. In this formulation, energy non conservation is tolerated and can source an effective cosmological constant \cite{josset2017}.} in a convincing way (which is arguably far from trivial) and if robust criteria can be found to make them less ad hoc (see \eg \cite{tilloy2017}), then further optimism will be warranted. Unless, of course, experiments provide a hint in the meantime \cite{bose2017}.

\begin{acknowledgments} I thank the participants of the 5th International Summer School in Philosophy of Physics in Saig, Germany, for pushing me to write down the model presented here. I am also grateful to D. Lazarovici for sharp comments and helpful suggestions. This work was made possible by support from the Alexander von Humboldt foundation and the Agence
Nationale de la Recherche (ANR) contract ANR-14-CE25-0003-01.
\end{acknowledgments}

\bibliographystyle{apsrev4-1}
\bibliography{main}

\begin{thebibliography}{25}%
\makeatletter
\providecommand \@ifxundefined [1]{%
 \@ifx{#1\undefined}
}%
\providecommand \@ifnum [1]{%
 \ifnum #1\expandafter \@firstoftwo
 \else \expandafter \@secondoftwo
 \fi
}%
\providecommand \@ifx [1]{%
 \ifx #1\expandafter \@firstoftwo
 \else \expandafter \@secondoftwo
 \fi
}%
\providecommand \natexlab [1]{#1}%
\providecommand \enquote  [1]{``#1''}%
\providecommand \bibnamefont  [1]{#1}%
\providecommand \bibfnamefont [1]{#1}%
\providecommand \citenamefont [1]{#1}%
\providecommand \href@noop [0]{\@secondoftwo}%
\providecommand \href [0]{\begingroup \@sanitize@url \@href}%
\providecommand \@href[1]{\@@startlink{#1}\@@href}%
\providecommand \@@href[1]{\endgroup#1\@@endlink}%
\providecommand \@sanitize@url [0]{\catcode `\\12\catcode `\$12\catcode
  `\&12\catcode `\#12\catcode `\^12\catcode `\_12\catcode `\%12\relax}%
\providecommand \@@startlink[1]{}%
\providecommand \@@endlink[0]{}%
\providecommand \url  [0]{\begingroup\@sanitize@url \@url }%
\providecommand \@url [1]{\endgroup\@href {#1}{\urlprefix }}%
\providecommand \urlprefix  [0]{URL }%
\providecommand \Eprint [0]{\href }%
\providecommand \doibase [0]{http://dx.doi.org/}%
\providecommand \selectlanguage [0]{\@gobble}%
\providecommand \bibinfo  [0]{\@secondoftwo}%
\providecommand \bibfield  [0]{\@secondoftwo}%
\providecommand \translation [1]{[#1]}%
\providecommand \BibitemOpen [0]{}%
\providecommand \bibitemStop [0]{}%
\providecommand \bibitemNoStop [0]{.\EOS\space}%
\providecommand \EOS [0]{\spacefactor3000\relax}%
\providecommand \BibitemShut  [1]{\csname bibitem#1\endcsname}%
\let\auto@bib@innerbib\@empty
\bibitem [{\citenamefont {Eppley}\ and\ \citenamefont
  {Hannah}(1977)}]{eppley1977}%
  \BibitemOpen
  \bibfield  {author} {\bibinfo {author} {\bibfnamefont {K.}~\bibnamefont
  {Eppley}}\ and\ \bibinfo {author} {\bibfnamefont {E.}~\bibnamefont
  {Hannah}},\ }\href {\doibase 10.1007/BF00715241} {\bibfield  {journal}
  {\bibinfo  {journal} {Found. Phys.}\ }\textbf {\bibinfo {volume} {7}},\
  \bibinfo {pages} {51} (\bibinfo {year} {1977})}\BibitemShut {NoStop}%
\bibitem [{\citenamefont {Albers}\ \emph {et~al.}(2008)\citenamefont {Albers},
  \citenamefont {Kiefer},\ and\ \citenamefont {Reginatto}}]{albers2008}%
  \BibitemOpen
  \bibfield  {author} {\bibinfo {author} {\bibfnamefont {M.}~\bibnamefont
  {Albers}}, \bibinfo {author} {\bibfnamefont {C.}~\bibnamefont {Kiefer}}, \
  and\ \bibinfo {author} {\bibfnamefont {M.}~\bibnamefont {Reginatto}},\ }\href
  {\doibase 10.1103/PhysRevD.78.064051} {\bibfield  {journal} {\bibinfo
  {journal} {Phys. Rev. D}\ }\textbf {\bibinfo {volume} {78}},\ \bibinfo
  {pages} {064051} (\bibinfo {year} {2008})}\BibitemShut {NoStop}%
\bibitem [{\citenamefont {Kafri}\ \emph {et~al.}(2014)\citenamefont {Kafri},
  \citenamefont {Taylor},\ and\ \citenamefont {Milburn}}]{kafri2014}%
  \BibitemOpen
  \bibfield  {author} {\bibinfo {author} {\bibfnamefont {D.}~\bibnamefont
  {Kafri}}, \bibinfo {author} {\bibfnamefont {J.~M.}\ \bibnamefont {Taylor}}, \
  and\ \bibinfo {author} {\bibfnamefont {G.~J.}\ \bibnamefont {Milburn}},\
  }\href {\doibase https://doi.org/10.1088/1367-2630/16/6/065020} {\bibfield
  {journal} {\bibinfo  {journal} {New J. Phys.}\ }\textbf {\bibinfo {volume}
  {16}},\ \bibinfo {pages} {065020} (\bibinfo {year} {2014})}\BibitemShut
  {NoStop}%
\bibitem [{\citenamefont {Kafri}\ \emph {et~al.}(2015)\citenamefont {Kafri},
  \citenamefont {Milburn},\ and\ \citenamefont {Taylor}}]{kafri2015}%
  \BibitemOpen
  \bibfield  {author} {\bibinfo {author} {\bibfnamefont {D.}~\bibnamefont
  {Kafri}}, \bibinfo {author} {\bibfnamefont {G.~J.}\ \bibnamefont {Milburn}},
  \ and\ \bibinfo {author} {\bibfnamefont {J.~M.}\ \bibnamefont {Taylor}},\
  }\href {\doibase https://doi.org/10.1088/1367-2630/17/1/015006} {\bibfield
  {journal} {\bibinfo  {journal} {New J. Phys.}\ }\textbf {\bibinfo {volume}
  {17}},\ \bibinfo {pages} {015006} (\bibinfo {year} {2015})}\BibitemShut
  {NoStop}%
\bibitem [{\citenamefont {Tilloy}\ and\ \citenamefont
  {Di\'osi}(2016)}]{tilloy2016}%
  \BibitemOpen
  \bibfield  {author} {\bibinfo {author} {\bibfnamefont {A.}~\bibnamefont
  {Tilloy}}\ and\ \bibinfo {author} {\bibfnamefont {L.}~\bibnamefont
  {Di\'osi}},\ }\href {\doibase 10.1103/PhysRevD.93.024026} {\bibfield
  {journal} {\bibinfo  {journal} {Phys. Rev. D}\ }\textbf {\bibinfo {volume}
  {93}},\ \bibinfo {pages} {024026} (\bibinfo {year} {2016})}\BibitemShut
  {NoStop}%
\bibitem [{\citenamefont {Tilloy}\ and\ \citenamefont
  {Di\'osi}(2017)}]{tilloy2017}%
  \BibitemOpen
  \bibfield  {author} {\bibinfo {author} {\bibfnamefont {A.}~\bibnamefont
  {Tilloy}}\ and\ \bibinfo {author} {\bibfnamefont {L.}~\bibnamefont
  {Di\'osi}},\ }\href {https://arxiv.org/abs/1706.01856} {\bibfield  {journal}
  {\bibinfo  {journal} {arXiv:1706.01856}\ } (\bibinfo {year}
  {2017})}\BibitemShut {NoStop}%
\bibitem [{\citenamefont {Ghirardi}\ \emph {et~al.}(1986)\citenamefont
  {Ghirardi}, \citenamefont {Rimini},\ and\ \citenamefont
  {Weber}}]{ghirardi1986}%
  \BibitemOpen
  \bibfield  {author} {\bibinfo {author} {\bibfnamefont {G.~C.}\ \bibnamefont
  {Ghirardi}}, \bibinfo {author} {\bibfnamefont {A.}~\bibnamefont {Rimini}}, \
  and\ \bibinfo {author} {\bibfnamefont {T.}~\bibnamefont {Weber}},\ }\href
  {\doibase 10.1103/PhysRevD.34.470} {\bibfield  {journal} {\bibinfo  {journal}
  {Phys. Rev. D}\ }\textbf {\bibinfo {volume} {34}},\ \bibinfo {pages} {470}
  (\bibinfo {year} {1986})}\BibitemShut {NoStop}%
\bibitem [{\citenamefont {Di\'osi}(1989)}]{diosi1989}%
  \BibitemOpen
  \bibfield  {author} {\bibinfo {author} {\bibfnamefont {L.}~\bibnamefont
  {Di\'osi}},\ }\href {\doibase 10.1103/PhysRevA.40.1165} {\bibfield  {journal}
  {\bibinfo  {journal} {Phys. Rev. A}\ }\textbf {\bibinfo {volume} {40}},\
  \bibinfo {pages} {1165} (\bibinfo {year} {1989})}\BibitemShut {NoStop}%
\bibitem [{\citenamefont {Ghirardi}\ \emph {et~al.}(1990)\citenamefont
  {Ghirardi}, \citenamefont {Pearle},\ and\ \citenamefont
  {Rimini}}]{ghirardi1990csl}%
  \BibitemOpen
  \bibfield  {author} {\bibinfo {author} {\bibfnamefont {G.~C.}\ \bibnamefont
  {Ghirardi}}, \bibinfo {author} {\bibfnamefont {P.}~\bibnamefont {Pearle}}, \
  and\ \bibinfo {author} {\bibfnamefont {A.}~\bibnamefont {Rimini}},\ }\href
  {\doibase 10.1103/PhysRevA.42.78} {\bibfield  {journal} {\bibinfo  {journal}
  {Phys. Rev. A}\ }\textbf {\bibinfo {volume} {42}},\ \bibinfo {pages} {78}
  (\bibinfo {year} {1990})}\BibitemShut {NoStop}%
\bibitem [{\citenamefont {Bassi}\ \emph {et~al.}(2013)\citenamefont {Bassi},
  \citenamefont {Lochan}, \citenamefont {Satin}, \citenamefont {Singh},\ and\
  \citenamefont {Ulbricht}}]{bassi2013}%
  \BibitemOpen
  \bibfield  {author} {\bibinfo {author} {\bibfnamefont {A.}~\bibnamefont
  {Bassi}}, \bibinfo {author} {\bibfnamefont {K.}~\bibnamefont {Lochan}},
  \bibinfo {author} {\bibfnamefont {S.}~\bibnamefont {Satin}}, \bibinfo
  {author} {\bibfnamefont {T.~P.}\ \bibnamefont {Singh}}, \ and\ \bibinfo
  {author} {\bibfnamefont {H.}~\bibnamefont {Ulbricht}},\ }\href {\doibase
  10.1103/RevModPhys.85.471} {\bibfield  {journal} {\bibinfo  {journal} {Rev.
  Mod. Phys.}\ }\textbf {\bibinfo {volume} {85}},\ \bibinfo {pages} {471}
  (\bibinfo {year} {2013})}\BibitemShut {NoStop}%
\bibitem [{\citenamefont {Feldmann}\ and\ \citenamefont
  {Tumulka}(2012)}]{feldmann2012}%
  \BibitemOpen
  \bibfield  {author} {\bibinfo {author} {\bibfnamefont {W.}~\bibnamefont
  {Feldmann}}\ and\ \bibinfo {author} {\bibfnamefont {R.}~\bibnamefont
  {Tumulka}},\ }\href {\doibase 10.1088/1751-8113/45/6/065304} {\bibfield
  {journal} {\bibinfo  {journal} {J. Phys. A: Math. Theor.}\ }\textbf {\bibinfo
  {volume} {45}},\ \bibinfo {pages} {065304} (\bibinfo {year}
  {2012})}\BibitemShut {NoStop}%
\bibitem [{\citenamefont {Bell}(1987)}]{bell1987}%
  \BibitemOpen
  \bibfield  {author} {\bibinfo {author} {\bibfnamefont {J.}~\bibnamefont
  {Bell}},\ }\enquote {\bibinfo {title} {Are there quantum jumps?}}\ in\ \href
  {\doibase 10.1017/CBO9780511564253.005} {\emph {\bibinfo {booktitle}
  {Schr\"odinger: Centenary Celebration of a Polymath}}},\ \bibinfo {editor}
  {edited by\ \bibinfo {editor} {\bibfnamefont {C.~W.}\ \bibnamefont
  {Kilmister}}}\ (\bibinfo  {publisher} {Cambridge University Press},\ \bibinfo
  {year} {1987})\ pp.\ \bibinfo {pages} {41--52}\BibitemShut {NoStop}%
\bibitem [{\citenamefont {Bell}(1976)}]{bell1976}%
  \BibitemOpen
  \bibfield  {author} {\bibinfo {author} {\bibfnamefont {J.~S.}\ \bibnamefont
  {Bell}},\ }\href {http://cds.cern.ch/record/980036/files/197508125.pdf}
  {\bibfield  {journal} {\bibinfo  {journal} {Epistemological Letters}\
  }\textbf {\bibinfo {volume} {9}} (\bibinfo {year} {1976})}\BibitemShut
  {NoStop}%
\bibitem [{\citenamefont {Allori}\ \emph {et~al.}(2014)\citenamefont {Allori},
  \citenamefont {Goldstein}, \citenamefont {Tumulka},\ and\ \citenamefont
  {Zangh{\`\i}}}]{allori2014}%
  \BibitemOpen
  \bibfield  {author} {\bibinfo {author} {\bibfnamefont {V.}~\bibnamefont
  {Allori}}, \bibinfo {author} {\bibfnamefont {S.}~\bibnamefont {Goldstein}},
  \bibinfo {author} {\bibfnamefont {R.}~\bibnamefont {Tumulka}}, \ and\
  \bibinfo {author} {\bibfnamefont {N.}~\bibnamefont {Zangh{\`\i}}},\ }\href
  {\doibase 10.1093/bjps/axs048} {\bibfield  {journal} {\bibinfo  {journal}
  {Brit. J. Philos. Sci.}\ }\textbf {\bibinfo {volume} {65}},\ \bibinfo {pages}
  {323} (\bibinfo {year} {2014})}\BibitemShut {NoStop}%
\bibitem [{\citenamefont {Allori}(2015)}]{allori2015}%
  \BibitemOpen
  \bibfield  {author} {\bibinfo {author} {\bibfnamefont {V.}~\bibnamefont
  {Allori}},\ }\href {http://www.ijqf.org/archives/2394} {\bibfield  {journal}
  {\bibinfo  {journal} {International Journal of Quantum Foundations}\ }\textbf
  {\bibinfo {volume} {1}},\ \bibinfo {pages} {107} (\bibinfo {year}
  {2015})}\BibitemShut {NoStop}%
\bibitem [{\citenamefont {Tumulka}(2011)}]{tumulka2011}%
  \BibitemOpen
  \bibfield  {author} {\bibinfo {author} {\bibfnamefont {R.}~\bibnamefont
  {Tumulka}},\ }\href {https://arxiv.org/abs/1102.5767} {\bibfield  {journal}
  {\bibinfo  {journal} {arXiv:1102.5767}\ } (\bibinfo {year}
  {2011})}\BibitemShut {NoStop}%
\bibitem [{\citenamefont {Gisin}(1990)}]{gisin1990}%
  \BibitemOpen
  \bibfield  {author} {\bibinfo {author} {\bibfnamefont {N.}~\bibnamefont
  {Gisin}},\ }\href {\doibase http://dx.doi.org/10.1016/0375-9601(90)90786-N}
  {\bibfield  {journal} {\bibinfo  {journal} {Phys. Lett. A}\ }\textbf
  {\bibinfo {volume} {143}},\ \bibinfo {pages} {1} (\bibinfo {year}
  {1990})}\BibitemShut {NoStop}%
\bibitem [{\citenamefont {Polchinski}(1991)}]{polchinski1991}%
  \BibitemOpen
  \bibfield  {author} {\bibinfo {author} {\bibfnamefont {J.}~\bibnamefont
  {Polchinski}},\ }\href {\doibase 10.1103/PhysRevLett.66.397} {\bibfield
  {journal} {\bibinfo  {journal} {Phys. Rev. Lett.}\ }\textbf {\bibinfo
  {volume} {66}},\ \bibinfo {pages} {397} (\bibinfo {year} {1991})}\BibitemShut
  {NoStop}%
\bibitem [{\citenamefont {Bassi}\ and\ \citenamefont
  {Hejazi}(2015)}]{bassi2015}%
  \BibitemOpen
  \bibfield  {author} {\bibinfo {author} {\bibfnamefont {A.}~\bibnamefont
  {Bassi}}\ and\ \bibinfo {author} {\bibfnamefont {K.}~\bibnamefont {Hejazi}},\
  }\href {\doibase 10.1088/0143-0807/36/5/055027} {\bibfield  {journal}
  {\bibinfo  {journal} {Eur. J. Phys.}\ }\textbf {\bibinfo {volume} {36}},\
  \bibinfo {pages} {055027} (\bibinfo {year} {2015})}\BibitemShut {NoStop}%
\bibitem [{\citenamefont {D{\"u}rr}\ and\ \citenamefont
  {Teufel}(2009)}]{durr2009}%
  \BibitemOpen
  \bibfield  {author} {\bibinfo {author} {\bibfnamefont {D.}~\bibnamefont
  {D{\"u}rr}}\ and\ \bibinfo {author} {\bibfnamefont {S.}~\bibnamefont
  {Teufel}},\ }\href@noop {} {\emph {\bibinfo {title} {Bohmian Mechanics}}}\
  (\bibinfo  {publisher} {Springer Berlin Heidelberg},\ \bibinfo {year}
  {2009})\BibitemShut {NoStop}%
\bibitem [{\citenamefont {{Derakhshani}}(2014)}]{derakhshani2014}%
  \BibitemOpen
  \bibfield  {author} {\bibinfo {author} {\bibfnamefont {M.}~\bibnamefont
  {{Derakhshani}}},\ }\href {\doibase
  http://dx.doi.org/10.1016/j.physleta.2014.02.005} {\bibfield  {journal}
  {\bibinfo  {journal} {Phys. Lett. A}\ }\textbf {\bibinfo {volume} {378}},\
  \bibinfo {pages} {990 } (\bibinfo {year} {2014})}\BibitemShut {NoStop}%
\bibitem [{\citenamefont {{Hagedorn}}(2015)}]{hagedorn2015}%
  \BibitemOpen
  \bibfield  {author} {\bibinfo {author} {\bibfnamefont {C.~A.}\ \bibnamefont
  {{Hagedorn}}},\ }\emph {\bibinfo {title} {{A Sub-Millimeter Parallel-Plate
  Test of Gravity}}},\ \href@noop {} {Ph.D. thesis},\ \bibinfo  {school}
  {University of Washington} (\bibinfo {year} {2015})\BibitemShut {NoStop}%
\bibitem [{\citenamefont {Bahrami}\ \emph {et~al.}(2014)\citenamefont
  {Bahrami}, \citenamefont {Gro{\ss}ardt}, \citenamefont {Donadi},\ and\
  \citenamefont {Bassi}}]{bahrami2014}%
  \BibitemOpen
  \bibfield  {author} {\bibinfo {author} {\bibfnamefont {M.}~\bibnamefont
  {Bahrami}}, \bibinfo {author} {\bibfnamefont {A.}~\bibnamefont
  {Gro{\ss}ardt}}, \bibinfo {author} {\bibfnamefont {S.}~\bibnamefont
  {Donadi}}, \ and\ \bibinfo {author} {\bibfnamefont {A.}~\bibnamefont
  {Bassi}},\ }\href {\doibase 10.1088/1367-2630/16/11/115007} {\bibfield
  {journal} {\bibinfo  {journal} {New J. Phys.}\ }\textbf {\bibinfo {volume}
  {16}},\ \bibinfo {pages} {115007} (\bibinfo {year} {2014})}\BibitemShut
  {NoStop}%
\bibitem [{\citenamefont {Josset}\ \emph {et~al.}(2017)\citenamefont {Josset},
  \citenamefont {Perez},\ and\ \citenamefont {Sudarsky}}]{josset2017}%
  \BibitemOpen
  \bibfield  {author} {\bibinfo {author} {\bibfnamefont {T.}~\bibnamefont
  {Josset}}, \bibinfo {author} {\bibfnamefont {A.}~\bibnamefont {Perez}}, \
  and\ \bibinfo {author} {\bibfnamefont {D.}~\bibnamefont {Sudarsky}},\ }\href
  {\doibase 10.1103/PhysRevLett.118.021102} {\bibfield  {journal} {\bibinfo
  {journal} {Phys. Rev. Lett.}\ }\textbf {\bibinfo {volume} {118}},\ \bibinfo
  {pages} {021102} (\bibinfo {year} {2017})}\BibitemShut {NoStop}%
\bibitem [{\citenamefont {Bose}\ \emph {et~al.}(2017)\citenamefont {Bose},
  \citenamefont {Mazumdar}, \citenamefont {Morley}, \citenamefont {Ulbricht},
  \citenamefont {Toros}, \citenamefont {Paternostro}, \citenamefont {Geraci},
  \citenamefont {Barker}, \citenamefont {Kim},\ and\ \citenamefont
  {Milburn}}]{bose2017}%
  \BibitemOpen
  \bibfield  {author} {\bibinfo {author} {\bibfnamefont {S.}~\bibnamefont
  {Bose}}, \bibinfo {author} {\bibfnamefont {A.}~\bibnamefont {Mazumdar}},
  \bibinfo {author} {\bibfnamefont {G.~W.}\ \bibnamefont {Morley}}, \bibinfo
  {author} {\bibfnamefont {H.}~\bibnamefont {Ulbricht}}, \bibinfo {author}
  {\bibfnamefont {M.}~\bibnamefont {Toros}}, \bibinfo {author} {\bibfnamefont
  {M.}~\bibnamefont {Paternostro}}, \bibinfo {author} {\bibfnamefont
  {A.}~\bibnamefont {Geraci}}, \bibinfo {author} {\bibfnamefont
  {P.}~\bibnamefont {Barker}}, \bibinfo {author} {\bibfnamefont {M.~S.}\
  \bibnamefont {Kim}}, \ and\ \bibinfo {author} {\bibfnamefont
  {G.}~\bibnamefont {Milburn}},\ }\href {https://arxiv.org/abs/1707.06050}
  {\bibfield  {journal} {\bibinfo  {journal} {arXiv:1707.06050}\ } (\bibinfo
  {year} {2017})}\BibitemShut {NoStop}%
\end{thebibliography}%

\end{document}